# Molecular engineering for crack-free transfer of single-crystalline oxide membranes onto silicon

Jeongkeun Song[1*], Jihye Choi[2,3*], Benjamin Doughty[3], Young-Hoon Kim[4], Keun-Yeol Park[5], Uvinduni I. Premadasa[3], Miaofang Chi[4,6], Seung Sae Hong[7], Celesta S. Chang[5,8], Zheng Gai[4], Ambrose Seo[9], Tomonori Saito[2,3], and Ho Nyung Lee[1]

[1]*Materials Science and Technology Division, Oak Ridge National Laboratory, Oak Ridge, TN 37831, USA*

[2]*Bredesen Center for Interdisciplinary Research and Education, University of Tennessee, Knoxville, TN 37966, USA*

[3]*Chemical Sciences Division, Oak Ridge National Laboratory, Oak Ridge, TN 37831, USA*

[4]*Center for Nanophase Materials Sciences, Oak Ridge National Laboratory, Oak Ridge, TN 37831, USA*

[5]*Department of Physics and Astronomy, Seoul National University, Seoul, 08826, Republic of Korea*

[6]*Department of Mechanical Engineering and Materials Science, Duke University, Durham, NC 27708, USA*

[7]*Department of Materials Science and Engineering, University of California, Davis, CA 95616, USA*

[8]*Institute of Applied Physics, Seoul National University, Seoul, 08826, Republic of Korea*

[9]*Department of Physics and Astronomy, University of Kentucky, Lexington, KY 40506, USA.*

Correspondence to: hnlee@ornl.gov

Notice: This manuscript has been authored by UT-Battelle, LLC, under contract DE-AC05-00OR22725 with the US Department of Energy (DOE). The US government retains and the publisher, by accepting the article for publication, acknowledges that the US government retains a nonexclusive, paid-up, irrevocable, worldwide license to publish or reproduce the published form of this manuscript, or allow others to do so, for US government purposes. DOE will provide public access to these results of federally sponsored research in accordance with the DOE Public Access Plan (http://energy.gov/downloads/doe-public-access-plan).

**Abstract**

Heterogeneous integration of single-crystalline complex oxides with silicon offers a route to incorporating their functionalities into next-generation microelectronics. Released from their growth substrates, freestanding oxide membranes enable this integration. However, cracks, wrinkles, and fragmentation during release and transfer limit membrane size and quality. Here, we present an interfacial energy-controlled release strategy that enables crack-free, single-crystalline freestanding oxide membranes with centimeter-scale lateral dimensions. Using $SrTiO_3$ as a model system, time-resolved release studies identify water-induced blistering at the membrane–support interface during prolonged water exposure as the origin of membrane cracking. Lowering the molecular weight of the poly(propylene carbonate) support through polymer deconstruction increases its polarity and surface energy, suppresses interfacial blistering, and reduces organic residue after support removal. The resulting $SrTiO_3$ membranes exhibit optical properties comparable to those of bulk single crystals. We extend this approach to the metallic ferromagnet $La_{0.7}Sr_{0.3}MnO_3$, obtaining crack-free membranes on Si that preserve ferromagnetism and metallicity.

## Introduction

Over the past decades, silicon complementary metal–oxide–semiconductor (CMOS) technology has advanced through the continuous scaling of transistors, allowing tens of billions of devices to be integrated on a single chip [1,2]. However, as transistor dimensions approach their physical limits, further scaling introduces challenges [3-5]. including quantum tunneling and heat dissipation. Heterogeneous integration of complex oxides on silicon is particularly attractive in this regard [6,7], as it can add new functionalities to silicon, such as ferroelectricity, magnetism, and superconductivity [8-10], which arise from the strong coupling among charge, spin, orbital, and lattice degrees of freedom [11,12]. Many efforts have been made toward this goal, including epitaxial growth on buffer layers and wafer bonding. Despite these efforts, the direct epitaxial growth of high-quality complex oxides on silicon remains challenging due to the oxidation of the Si surface and the defects caused by lattice and thermal expansion mismatches [13-15].

Conventional thin-film epitaxy has been a powerful means of controlling the free energy of complex oxides through substrate-imposed boundary conditions [16-18]. At the same time, these conditions restrict the accessible material combinations and structural configurations [19,20]. Recent advances in lift-off techniques for complex oxides offer a promising route to overcome these limitations [6,19,21-24]. Through either selective etching of a sacrificial layer or mechanical exfoliation, the oxide film is decoupled from its growth substrate. The released membrane can then be integrated with substrates and materials that are incompatible with direct growth, such as Si, flexible substrates, and two-dimensional (2D) materials. Moreover, freestanding membranes can be translated, rotated, and stacked after growth, introducing additional geometrical degrees of freedom such as interfacial orientation and twist angle. These new degrees of freedom have opened opportunities to explore emergent electronic and magnetic phenomena, including polar textures [25-29], moiré-driven magnetism [30-32], flat bands [19,32,33], and superconductivity [34-36].

To fully exploit these opportunities, scalable fabrication of freestanding membranes with preserved structural integrity is essential. However, microcracks, wrinkles, and fragmentation are often introduced during release and transfer, limiting both the size and quality of the membranes [22-24,37]. Considerable effort has focused on improving sacrificial-layer design, lattice matching, and dissolution kinetics [38-41], as well as the mechanical properties of polymer supports to compensate the strain energy released upon membrane lift-off [21]. Yet strain release may not be the sole origin of cracking. For instance, unlike van der Waals materials, which are often transferred by dry methods [42], oxide membranes that are released through a wet process expose the polymer support directly to the etchant or water. This exposure can change the characteristics of the polymer support and its interface with the membrane, affecting the structural integrity and scalability of the membranes. Nevertheless, how the polymer support interacts with the etchant and how this interaction affects the membrane remain poorly understood.

In this study, we establish membrane–support interfacial energy engineering as a strategy for the scalable integration of crack-free, single-crystalline oxide membranes on Si. Using time-resolved water-exposure experiments, we identify interfacial blistering as the primary origin of transfer-induced cracking. By controlling the molecular weight of poly(propylene carbonate) (PPC) through glycolysis, we show that lowering the molecular weight increases the surface polarity and surface energy of the support, which tunes the support interfacial energy, suppresses blister nucleation and growth even under prolonged water exposure, and reduces organic residue after support removal. As a result, we obtain large-scale, crack-free freestanding $SrTiO_3$ (STO) membranes with preserved crystallinity and optical properties. The same approach offer crack-free metallic ferromagnet, such as $La_{0.7}Sr_{0.3}MnO_3$, membranes on Si substrates that retain their ferromagnetic and metallic properties, demonstrating its applicability beyond STO. We believe that tailoring the membrane–support interface will provide a general route toward integrating high-quality freestanding complex oxides into heterogeneous, stacked, and twisted oxide architectures on Si.

**Results**

To investigate how the interaction between water and the polymer support influences the oxide membrane, we used STO membranes as a model system. STO/SAO-427 heterostructures ($Sr_4Al_2O_7$; orthorhombic, $a = 10.798$ Å, $b = 11.238$ Å, and $c = 25.732$ Å [43], corresponding to a pseudo-tetragonal unit cell with $a^* = b^* = 3.896$ Å and $c^* = 4.293$ Å [23]) were epitaxially grown on $TiO_2$-terminated (001) STO substrates by pulsed laser deposition (PLD) (Fig. 1A). A PPC support layer was then applied to the heterostructures (see Methods), and the SAO-427 sacrificial layer was etched in deionized water at room temperature within approximately 30 min [23]. The released membrane was subsequently transferred onto a target substrate, followed by thermal removal of the PPC support layer.

To examine the effect of water exposure, the released PPC/STO was kept in deionized water for different durations before transfer, measured from the start of immersion. After 30 min, the PPC/STO interface remained uniform without visible defects, and the transferred STO membrane remained intact (Fig. 1B). Interestingly, after 60 min, localized blisters emerged at the PPC/STO interface, coinciding with the first appearance of cracks after transfer (Fig. 1C). After 300 min, the blisters grew and their density increased substantially, leading to membrane fragmentation during transfer (Fig. 1D). This correlated evolution suggests that interfacial blistering acts as a precursor to transfer-induced cracking.

Blister formation is attributed to water permeation through the support layer and osmotic pressure-driven delamination at the membrane–support interface [44-46]. As water permeates through the polymer support, it accumulates at the PPC/STO interface, leading to blister nucleation and growth (Fig. 1 B to D). The blisters locally deform the membrane, and the resulting strain fractures the membrane, producing cracks that remain after transfer and PPC removal. Although faster dissolution of the sacrificial layer can shorten water exposure and thus limit blister formation, many oxide-membrane fabrication processes require prolonged aqueous etching [24,41,47]. For instance, complete dissolution of $Sr_3Al_2O_6$ (SAO-326), one of the most widely used sacrificial layers, takes several hours in deionized water at room temperature. In this sense, membrane–support interfacial stability is a key parameter for suppressing transfer-induced damage and achieving scalable fabrication of freestanding oxide membranes.

To understand the support-layer properties governing interfacial blistering, we first investigated the molecular and physical properties of PPC used for membrane transfer. As-received PPC exhibited substantial batch-to-batch variations in molecular weight and dispersity (fig. S1), together with residual cyclic propylene carbonate (cPC), which acts as a plasticizer (fig. S2). Solvent washing removed cPC and other low-molecular-weight species, increasing the number-average molecular weight ($M_n$) from approximately 119 to 183 kDa and reducing the dispersity ($Đ$) from 2.8 to 2.0 (fig. S1). We hereafter refer to the washed PPC as purified PPC.

To tune the molecular weight, as-received PPC was partially depolymerized through mTBD-catalyzed glycolysis [48,49] and then subjected to the same washing step, yielding a controlled series with molecular weight ranging from approximately 116 to 15 kDa (Fig. 2A and fig. S3). Molecular-weight reduction systematically modified the thermal, mechanical, and interfacial properties of the support layer. The glass-transition temperature decreased from approximately 36 to 27 °C with decreasing molecular weight, while PPC remained glassy under ambient conditions and retained its mechanical support function (Fig. 2B and fig. S4). The decomposition temperature also decreased from approximately 270 to 250 °C, indicating more facile thermal removal of the lower-molecular-weight support. The Young's modulus likewise decreased, from approximately 2.2 GPa for purified PPC to approximately 0.8 GPa at $M_n \approx 15$ kDa (Fig. 2C and fig. S5). In contrast, the interfacial properties evolved in the opposite direction. The water contact angle decreased from 98.0 to 77.7°, while the total surface energy increased from 27.7 to 34.4 mJ/m² as the molecular weight was reduced (Fig. 2D, fig. S6, and table S1). The nearly unchanged hexadecane contact angle indicates that this increase arises primarily from an enhanced polar component, which can be attributed to the higher density of hydroxyl end groups generated by glycolysis at lower molecular weight. These changes are expected to increase the affinity of the support toward the hydrophilic STO surface, suggesting that the membrane–support interfacial energy can be tuned through molecular-weight engineering.

To investigate how these competing mechanical and interfacial changes influence blister formation during prolonged water exposure, PPC/STO samples with different PPC molecular weights were immersed in deionized water at room temperature for 300 min (Fig. 2E). Both blister size and density progressively decreased with decreasing molecular weight, with pronounced suppression for $M_n \leq 57$ kDa. Notably, this onset coincides with the increase in wettability and polar surface energy (Fig. 2D and fig. S6). Particularly, the water contact angle drops 97.5 to 86.3° and the total surface energy increases from 27.7 to 30.8 mJ/m$^2$ (Fig. 2D and table S1), whereas the Young's modulus remains nearly unchanged (Fig. 2C). Moreover, blister suppression persists down to 15 kDa despite the substantial reduction in modulus (Fig. 2C and fig. S5), which would otherwise be expected to favor deformation and blister growth. These observations cannot be explained by the mechanical properties of the support alone. Instead, this behavior can be understood from the competition among osmotic pressure, support elasticity, and interfacial adhesion during wet release: the increased surface polarity strengthens interactions with the hydrophilic oxide surface and raises the resistance to interfacial delamination, outweighing the effect of reduced stiffness. Consequently, within the investigated molecular-weight range, membrane-support interfacial energy, rather than support stiffness, emerges as the dominant factor suppressing blister nucleation and growth.

After transfer, the polymer support must be removed. This is typically achieved by thermal decomposition (see Methods), which can leave residual polymer on the membrane surface owing to incomplete decomposition [42]. Since this temperature is close to or below the decomposition temperature of high-molecular-weight PPC (Fig. 2B), the lower decomposition temperature of the molecular-weight-engineered PPC is expected to enable more complete removal. We therefore examined whether molecular-weight engineering could also reduce organic residue after support removal. Sum-frequency generation (SFG) vibrational spectroscopy was performed on PPC-coated sapphire substrates subjected to the same thermal-removal protocol. As shown in Fig. 2F, the characteristic C–H vibrational intensity systematically decreases with decreasing molecular weight, indicating a smaller amount of residual organic species after thermal treatment.

Since SFG intensity depends on both the population and orientation of molecules, we further analyzed the symmetric (ss) and asymmetric (as) $CH_3$ stretching modes in the SSP and PPP polarization combinations [50,51]. The extracted orientational parameters show only minor variations among the samples (tables S2 and S3), suggesting that the decrease in SFG intensity mainly reflects reduced surface residue rather than changes in molecular orientation. Together, these results show that molecular-weight engineering of the support not only suppresses blister-induced membrane damage during wet release but also reduces organic residue after thermal removal, providing a basis for the crack-free, clean integration of freestanding oxide membranes demonstrated below.

Having established how molecular-weight engineering modifies the properties of the polymer support, we next examined how blister suppression affects the quality of the transferred membranes. Figure 3A displays an STO membrane transferred onto Si using as-received PPC after 200 min of water exposure. Although the membrane appears continuous at the macroscopic scale, optical microscopy reveals numerous cracks and voids across the membrane, consistent with water-induced interfacial blistering. In contrast, the STO membrane transferred using 15 kDa PPC under the same release conditions remains uniform and free of visible cracks over the entire membrane area, reaching up to the centimeter scale (Fig. 3B and fig. S7). This result indicates that blister suppression enables large-area, crack-free STO membranes even with slowly dissolving sacrificial layers.

X-ray diffraction (XRD) measurements of STO membranes transferred onto Si substrates showed only the STO 001 and 002 reflections, without detectable peaks associated with the sacrificial layer (Fig. 3C), confirming complete removal of the sacrificial layer and preservation of crystallinity following release and transfer. Although the Laue oscillations observed in the as-grown heterostructure became less pronounced after membrane release and transfer, the transferred membrane retained high crystalline quality. XRD rocking-curve measurements yielded a full width at half maximum (FWHM) of approximately 0.1° (Fig. 3D), indicating low mosaicity. Low-angle X-ray reflectivity (XRR) measurements further confirmed the thickness uniformity of the transferred membrane (Fig. 3E). Quantitative fitting using a multilayer model yielded an effective membrane thickness of approximately 13.0 nm and an interface roughness of approximately 2.9 Å.

To further assess the structural integrity of the transferred STO membranes, we performed high-angle annular dark-field scanning transmission electron microscopy (HAADF-STEM) in both cross-sectional and plane-view geometries. Figs. 3F and 3G show low- and high-magnification cross-sectional HAADF-STEM images of STO membranes transferred onto Si substrates. The membrane exhibits nearly parallel top and bottom interfaces together with uniform lattice contrast throughout the film thickness, indicating a flat morphology without discernible wrinkling, bending, or other transfer-induced deformation. The atomic-resolution plane-view HAADF-STEM image in Fig. 3H further resolves the perovskite lattice without observable lattice disorder or structural degradation. These results indicate that the molecular-weight-engineered transfer preserves the structural integrity of STO membranes from the macroscopic scale down to the atomic scale.

To investigate whether the intrinsic properties of the transferred membranes were preserved, we performed spectroscopic ellipsometry on STO membranes transferred onto $Al_2O_3$ substrates and extracted their dielectric function. STO is a wide-bandgap oxide (~3.2 eV) with strong ultraviolet absorption associated with O-2$p$ to Ti-3$d$ interband transitions. The extracted complex dielectric function of the transferred STO membrane closely matches that of single-crystal STO (Fig. 4A to C), exhibiting a sharp absorption edge ($E_g \sim 3.2$ eV) without detectable sub-bandgap absorption. These observations indicate that the intrinsic electronic structure of STO is preserved, with no evident defect-related optical signatures after release and transfer. Furthermore, we extended this molecular-weight-engineered transfer strategy to the perovskite ferromagnetic oxide metal $La_{0.7}Sr_{0.3}MnO_3$ (LSMO). The LSMO membrane transferred onto a Si substrate showed no visible cracks or physical defects over the millimeter-scale area (Fig. 4D), while XRD scans showed only the 00$l$ reflections of LSMO after transfer, confirming that the membrane retains its (001) orientation, and XRR yielded a membrane thickness of ~30 nm (fig. S8). Importantly, the transferred LSMO membrane preserves its magnetic and electronic transport properties. Magnetic hysteresis loops measured from 5 to 350 K reveal room-temperature ferromagnetism, with a magnetic moment of approximately 3.4 $\mu_B$/Mn at 5 K (Fig. 4E), comparable to the bulk value of approximately 3.5-3.6 $\mu_B$/Mn[52]. Temperature-dependent resistivity measurements showed metallic transport at low temperatures, followed by a resistivity maximum at 363 K (Fig. 4F). This resistivity maximum is consistent with enhanced spin-disorder scattering near the ferromagnetic transition and a change in charge transport associated with the double-exchange interaction in LSMO. These results demonstrate the general applicability of membrane–support interface engineering for the scalable fabrication of high-quality freestanding complex-oxide membranes.

## DISCUSSION

In summary, we establish membrane–support interfacial energy as a key design parameter for achieving large-area, crack-free freestanding oxide membranes. Time-resolved water-exposure experiments identify water-induced interfacial blistering as the primary origin of membrane damage during wet release and transfer. By tuning the membrane–support interfacial energy through the molecular weight of the PPC support, we show that enhanced interfacial affinity suppresses blister nucleation and growth, even though the mechanical stiffness of the support decreases. These results indicate that interfacial stability, rather than mechanical support alone, governs membrane integrity during aqueous release.

In addition, the molecular-weight-engineering strategy reduces organic residue after support removal, preserving both membrane integrity and surface cleanliness throughout the integration process. Using this approach, we obtain centimeter-scale, crack-free STO membranes on Si with high crystallinity and transferred STO membranes exhibit optical properties closely matching those of bulk single crystals. The strategy also extends to LSMO, yielding crack-free membranes on Si that retain their ferromagnetic and metallic properties, which demonstrates its applicability to functional complex oxides beyond STO. More broadly, control of the membrane–support interface provides a scalable route toward high-quality freestanding complex oxides for heterogeneous integration, including with silicon, and stacked or twisted oxide architectures.

## Methods

**Growth of epitaxial oxide heterostructures.** Epitaxial STO/SAO-427 heterostructures were grown on (001) STO substrates by pulsed laser deposition. Prior to film growth, (001) STO single-crystal substrates (Crystec GmbH) were etched in buffered oxide etchant and annealed at 1100 °C to achieve a $TiO_2$-terminated surface with well-defined step-terrace features. All depositions were carried out using a KrF excimer laser (Coherent LPXpro 305F, $\lambda$ = 248 nm) with a laser fluence fixed at 1.0 $Jcm^{-2}$. SAO-427 (20 nm in thickness) were grown using a polycrystalline $Sr_3Al_2O_6$ target at an oxygen partial pressure of 15 mTorr and a substrate temperature of 750 °C. Single-crystalline STO and polycrystalline $Sr_3Al_2O_6$ targets were employed. Subsequently, 13 nm STO layers were epitaxially grown on the sacrificial layers at an oxygen partial pressure of 15 mTorr and a substrate temperature of 700 °C. Real-time reflection high-energy electron diffraction (RHEED) was used to monitor the film growth process.

**Release and transfer of membranes.** To mechanically support the STO membrane during selective etching of the SAO-427 sacrificial layer in water, 2 μL of poly(propylene carbonate) (PPC; Sigma-Aldrich) solution (10 g of PPC per 100 mL of anisole) was drop-cast onto the epitaxial STO/ SAO-427 heterostructure. The PPC layer was cured on a hot plate at 70 °C for 10 min. The sample was then immersed in deionized water at room temperature, where the sacrificial layer was dissolved. Upon dissolution of the sacrificial layer, the PPC-supported STO membrane was released from the substrate and subsequently transferred onto $SiO_2$/Si substrates. In the main text, $SiO_2$/Si substrates are referred to as Si substrates for simplicity. After transfer, the PPC support layer was removed by thermal decomposition in air at 250 °C for 1 h, yielding freestanding STO membranes.

**Structural characterization.** High-resolution four-circle X-ray diffraction (X'Pert Pro, PANalytical; Cu $K\alpha_1$ radiation) was used to characterize the structural properties of the crystalline films. To investigate the atomic structure of freestanding membranes, HAADF-STEM imaging was performed in both cross-sectional and plane-view geometries. For cross-sectional imaging, STO membranes transferred onto Si substrates were first coated with electron-beam-induced carbon deposition, followed by focused ion beam specimen preparation using a Helios G4 FIB system (Thermo Fisher Scientific). Cross-sectional images were acquired using a Themis Z aberration-corrected and monochromated microscope (Thermo Fisher Scientific) operated at 300 kV, with a semi-convergence angle of 18 mrad and HAADF collection angles of 50–200 mrad. For plane-view imaging, a membrane was transferred to a TEM grid with 2 μm holes supported by a 100 nm-thick silicon nitride membrane (ESPER Biosciences). HAADF-STEM was performed using a NEOARM200CF microscope (JEOL Ltd.) operated at 200 kV, with a semi-convergence angle of 28 mrad and a HAADF detector angle range of 68–280 mrad.

**Electrical and magnetic characterization.** Electrical transport properties were measured via a standard four-point probe method using a commercial physical property measurement system (Dynacool; Quantum Design Inc.), which has a base temperature of 2 K and maximum magnetic field of 14 T. Au layer (thickness: ~50 nm) was deposited as an electrode by RF magnetron sputtering on LSMO membranes. The wiring between the sample holder and film was done by an ultrasonic wire bonder. A current of 2 mA was applied during the transport measurements. Magnetic measurements were performed using a Quantum Design Magnetic Property Measurement System (MPMS) equipped with a Vibrating Sample Magnetometer (VSM). The magnetic field was applied perpendicular to the sample surface. Field-dependent magnetization measurements were conducted over the range of –7 to +7 T. The diamagnetic background of the substrate was subtracted.

**PPC deconstruction assisted by mTBD.** PPC pellets were purchased from Sigma-Aldrich. PPC pellets (10 g) were mixed with ethylene glycol (EG, 3 equivalents relative to the PPC repeating unit) and dissolved in tetrahydrofuran (THF, 350 mL). The reaction mixture was maintained at 25 °C, and methyl-1,5,7-triazabicyclo[4.4.0]dec-5-ene (mTBD, 5 mol% relative to the PPC repeating unit) was added. After the desired reaction time, acetic acid (10 equivalents relative to mTBD) was introduced to deactivate the catalyst via acid-base neutralization. The solvent was removed under reduced pressure, and the resulting polymer was precipitated into deionized water. The precipitated polymer was washed with dichloromethane (DCM) and methanol at least three times and dried in a vacuum oven overnight.

**Molecular weight analysis by chromatography.** Size-exclusion chromatography (SEC) analysis was performed using an Agilent 1260 Infinity II LC system equipped with three PolyPore (250 × 4.6 $mm^2$) columns in series and a PolyPore (50 × 4.6 $mm^2$) guard column. THF was used as the mobile phase at a flow rate of 0.3 mL $min^{-1}$. Calibration was conducted using Agilent EasiVial PMMA standards with four discrete molecular weights. Detection was carried out using a differential refractive index (RI) detector and a light scattering detector operating with two angles at 90° and 15°. Number-average molecular weights ($M_n$), weight-average molecular weights ($M_w$), and dispersity ($Đ = M_w/M_n$) were determined using the Agilent GPC/SEC software, assuming complete mass elution.

**Thermogravimetric analysis.** Thermogravimetric analysis (TGA) was performed using a TGA55 (TA Instruments, Delaware, USA). Approximately 10 mg of sample was first heated to 100 °C at a rate of 5.0 °C min$^{-1}$ and cooled to room temperature under nitrogen flow for drying. The sample was subsequently heated from room temperature to 400 °C under air flow.

**Differential scanning calorimetry.** Differential scanning calorimetry (DSC) was performed using TA Instruments Q1000 DSC (Delaware, USA). Approximately 3 mg of sample was loaded into a sealed standard aluminum pan and heated at a rate of 10 °C min$^{-1}$ under nitrogen. The glass transition temperature ($T_g$) was determined from the point of inflection of the second heat flow curve.

**Tensile analysis.** Tensile stress-strain measurements of pristine and deconstructed PPC films were obtained using an Instron 3343 Universal Testing System (Massachusetts, USA) equipped with a 1 kN load cell, following ASTM D1708. Films were prepared by hot pressing at 100 °C for 1 min and subsequently punched to cut the dog-bone shaped films with a total length of 18 mm, a gauge width of 2.5 mm, and an average thickness of 0.5 mm. Mechanical testing was conducted at 22 °C. Samples were elongated at a rate of 100 mm min$^{-1}$ until failure or 1 mm min$^{-1}$ until failure. Young's modulus and yield strength were obtained from stress-strain data using the Instron Bluehill software.

**Static contact angle measurement.** Static contact angle measurements using polar (water) and nonpolar (hexadecane) solvents were performed with a ramé-hart contact angle goniometer to characterize the surface wettability of STO and PPC films with different molecular weights. Water droplets with volumes of 8.0 μL for STO and PPC were dispensed onto the sample surface, and the contact angles were measured immediately after droplet deposition. Hexadecane droplets with volume of 5.0 μL for PPC were dispensed onto PPC surfaces, and the contact angles were measured immediately after droplet deposition. Measurements were performed at three different locations on each sample, and the average values and standard deviations were calculated. The surface energies of PPC samples with different molecular weights were calculated using the Owens–Wendt equation[53,54]:

$$\gamma_l(1+\cos\theta) = 2\left(\sqrt{\gamma_s^d\gamma_l^d} + \sqrt{\gamma_s^p\gamma_l^p}\right) \tag{2}$$

where $\gamma_l$ is the total surface energy of the liquid, $\theta$ is the measured contact angle, $\gamma_s^d$ and $\gamma_s^p$ are the dispersive and polar components of the solid surface energy, and $\gamma_l^d$ and $\gamma_l^p$ are the corresponding dispersive and polar components of the liquid surface energy.

**Sum frequency generation (SFG) vibrational spectroscopy.** Details of the homebuilt SFG spectrometer used in experiments can be found elsewhere [55,56]. Of relevance to experiments performed here, broadband mid-infrared pulses centered at ~2900 $cm^{-1}$ were collinearly combined with near-infrared, spectrally narrowed pulses [57,58], and focused onto the sample PPC-sapphire interface at ~60° angle. The radiated SFG signal was collected in reflection, polarization resolved, spectrally filtered and dispersed in a spectrometer (Andor Kymera 328*i*) equipped with an electron-multiplying CCD camera (Andor, Newton) operating in a conventional collection mode with a 120-second exposure time. Replicate spectra were background subtracted before scaling by the nonresonant response of a gold film to correct for the spectral shape of the mid-IR pulse. Measurements were made in the SSP and PPP polarization combinations for orientational analysis.[50] SFG spectra were fit to a coherent summation of Lorentzians interfering with a nonresonant response ($\chi_{NR}^{(2)}$ with phase, $\phi$):

$$I_{SFG} \propto \left| \sum_q \frac{A_q}{\omega_{IR} - \omega_q + i\Gamma_q} + \chi_{NR}^{(2)} e^{i\phi} \right|^2 \tag{3}$$

where $I_{SFG}$ is the measured SFG intensity, $A_q$, $\Gamma_q$, and $\omega_q$, are mode specific amplitudes, widths and frequencies. A summary of fitting parameters is given in the Tables S2 and S3. The ratio of $A_q/\Gamma_q$ was used to quantify relative populations of residual PPC after thermal treatment.

**Ellipsometry measurement.** Spectroscopic ellipsometry was performed using an M-2000 ellipsometer (J.A. Woollam Co.) to obtain the real ($\varepsilon_1$) and imaginary ($\varepsilon_2$) parts of the dielectric function. Prior to all measurements, the instrument was calibrated at 300 K using a 25 nm $SiO_2$/Si wafer to minimize the mean square error (MSE) of the optical signal. Ellipsometric spectra were acquired at an incident angle of 70°, yielding $\Psi$ and $\Delta$ over the photon-energy range of 0.74 – 6.46 eV.

## Acknowledgments

This work was primarily sponsored by the U.S. Department of Energy (DOE), Office of Science, Basic Energy Sciences, Materials Sciences and Engineering Division (Oxide synthesis, Polymer deconstruction and characterization, Plane-view microscopy, X-ray and Optical measurements). SFG and part of contact-angle measurements were supported by the U.S. Department of Energy, Office of Science, Basic Energy Sciences, Chemical Sciences, Geosciences, and Biosciences Division. Plane-view microscopy work and magnetic characterization were conducted as part of a user project at the Center for Nanophase Materials Sciences (CNMS), which is a U.S. Department of Energy, Office of Science User Facility at Oak Ridge National Laboratory (ORNL). Cross-sectional microscopy work was supported by the National Research Foundation of Korea (NRF) grant funded by the Korean government (MSIT) (No. RS-2021-NR060087). Part of optical spectroscopy and data analysis was supported by the National Science Foundation Grant No. DMR-2426874.

## Author contributions

J.S., J.C., and H.N.L. conceived the idea and designed the experiments. J.S. and H.N.L. performed freestanding membrane fabrication, characterization, and data analysis. J.C. and T.S. performed polymer deconstruction, characterization, and data analysis. J.C. and U.I.P performed contact angle measurements. Y.-H. K. M. C. K.-Y. P. and C.S.C performed STEM measurements. B.D. performed SFG measurements. S.S.H. guided for the initial membrane transfer. J.S. and A.S. performed ellipsometric optical spectroscopy. J.S., J.C., and H.N.L. wrote the manuscript with contributions from all authors.

## Additional information

Supplementary information is available in the online version of the paper. Correspondence and requests for materials should be addressed to H. N. L.

## Competing financial interests

The authors declare no competing financial interests.

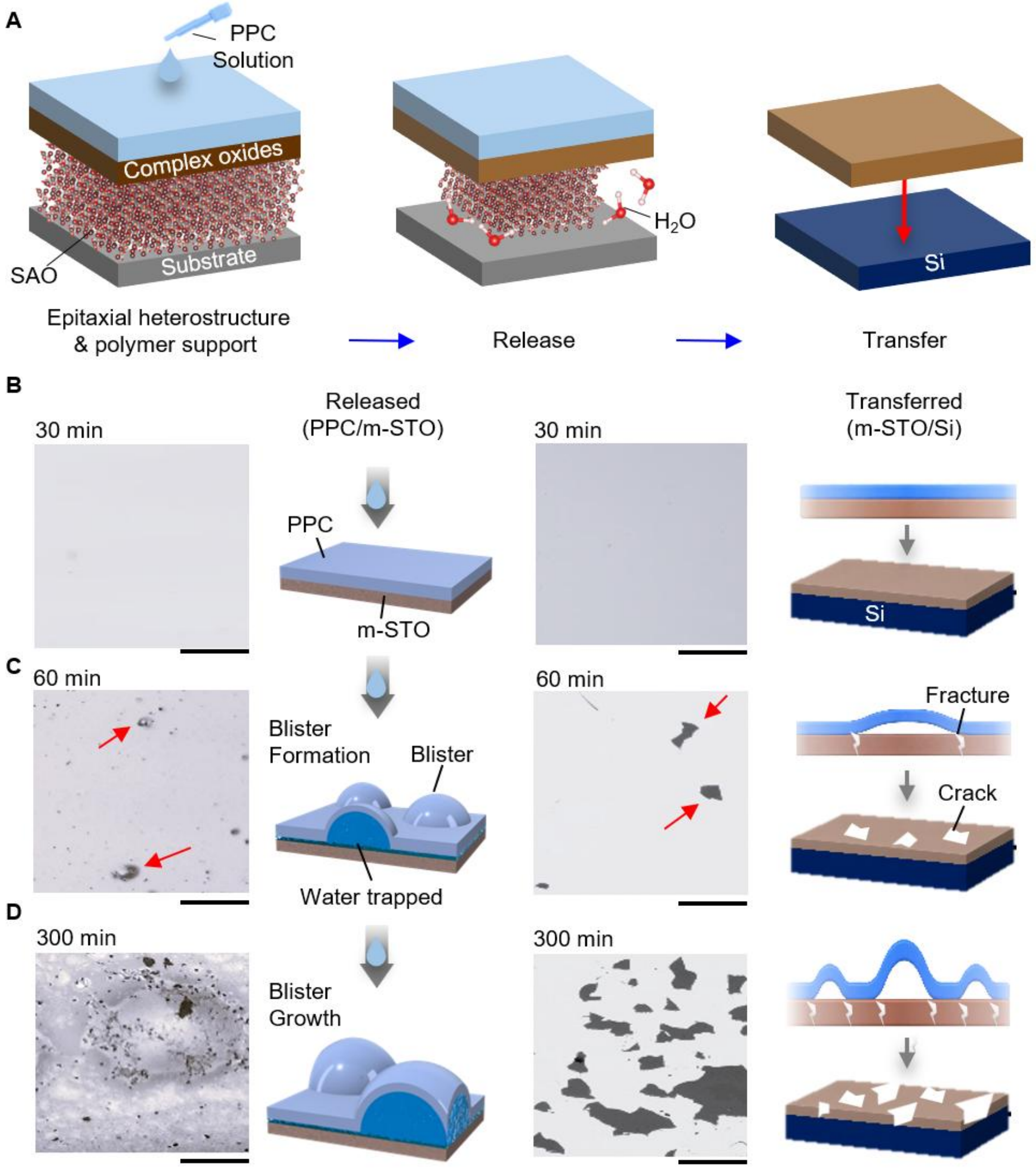

**Fig. 1. Water-exposure-dependent blister formation and structural degradation of freestanding oxide membranes. (A)** Schematic illustration of the fabrication and transfer process for freestanding $SrTiO_3$ (STO) membranes. A poly(propylene carbonate) (PPC) support layer is deposited onto an epitaxial STO/$Sr_4Al_2O_7$ (SAO) heterostructure. Selective dissolution of the water-soluble sacrificial layer in deionized water releases the PPC-supported STO membrane, which is subsequently transferred onto a $SiO_2$/Si substrate. For simplicity, $SiO_2$/Si substrates are referred to as Si substrates. **(B-D)** Evolution of PPC-supported STO membranes and corresponding transferred membranes with increasing water-exposure time. After 30 min of immersion, the membrane remains intact with no apparent macroscopic damage. After 60 min, localized blisters emerge in the PPC-supported membrane, accompanied by fracture and crack formation after transfer. Prolonged immersion for 300 min results in substantial blister growth and extensive fragmentation of the transferred STO membrane. Scale bars, 100 μm.

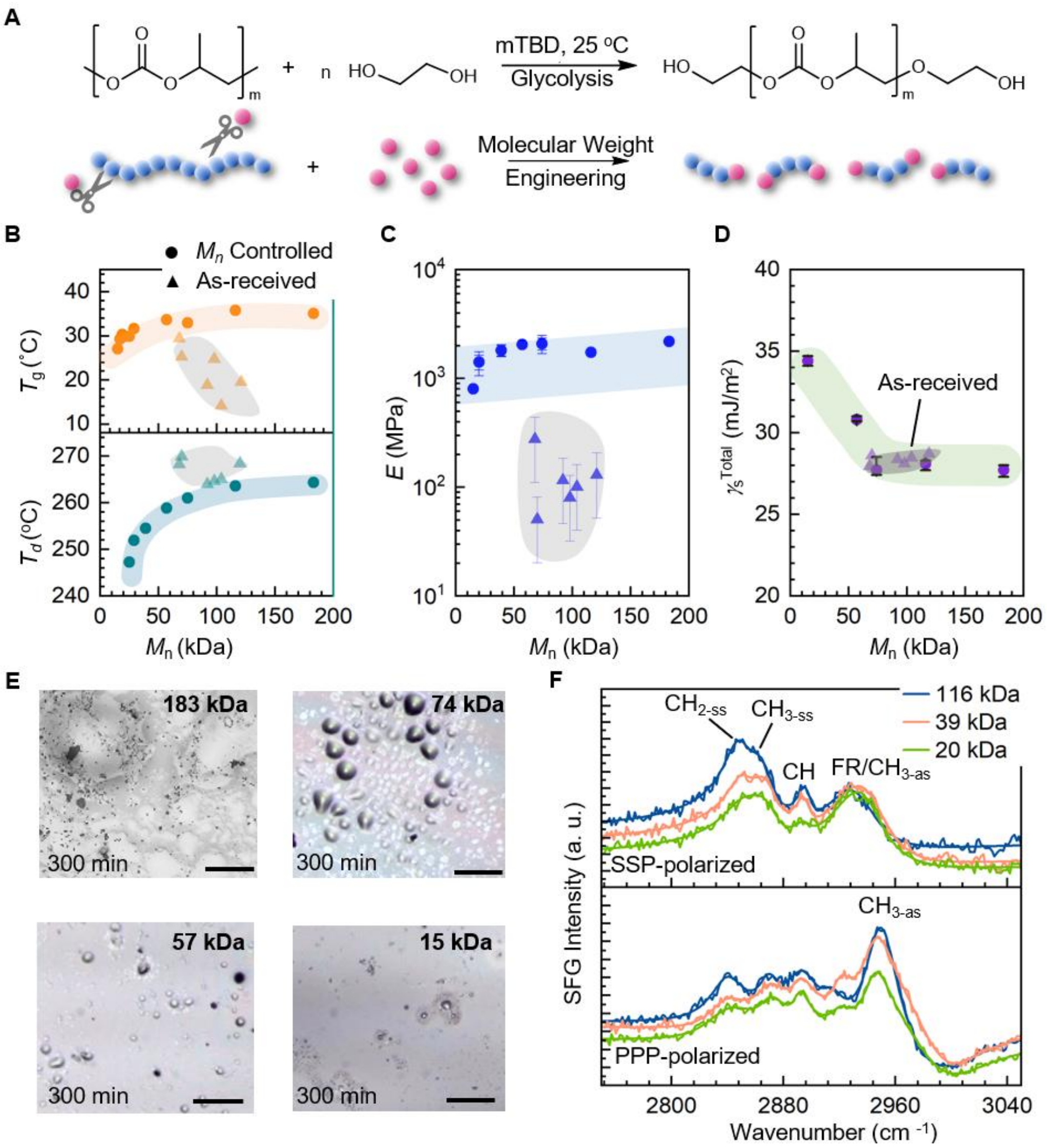

**Fig. 2. Suppression of blister formation during membrane release and transfer by molecular-weight engineering of PPC. (A)** Schematic illustration of mTBD-catalyzed glycolysis of poly(propylene carbonate) (PPC), showing controlled polymer deconstruction. **(B)** Glass transition temperature ($T_g$) and decomposition temperature ($T_d$) of pristine and deconstructed PPC as a function of number-average molecular weight ($M_n$), determined by DSC and TGA, respectively. **(C)** Young's modulus ($E$) of PPC extracted from tensile stress–strain measurements, indicating a reduction in stiffness at low molecular weight while remaining significantly higher than that of as-received PPC. **(D)** Total surface energy ($\gamma$) of PPC obtained from contact angle measurements using the Owens-Wendt method, showing an increase in surface energy with decreasing $M_n$. **(E)** Optical microscopy images of PPC-supported STO membranes after 300 min exposure to deionized water at room temperature. Scale bars, 100 µm. **(F)** SFG spectra of PPC films with different $M_n$ after thermal removal, collected in the SSP (top) and PPP (bottom) polarization configurations. The reduced SFG intensity for lower-molecular-weight PPC reflects a decrease in residual polymer after thermal removal (table S2 and S3).

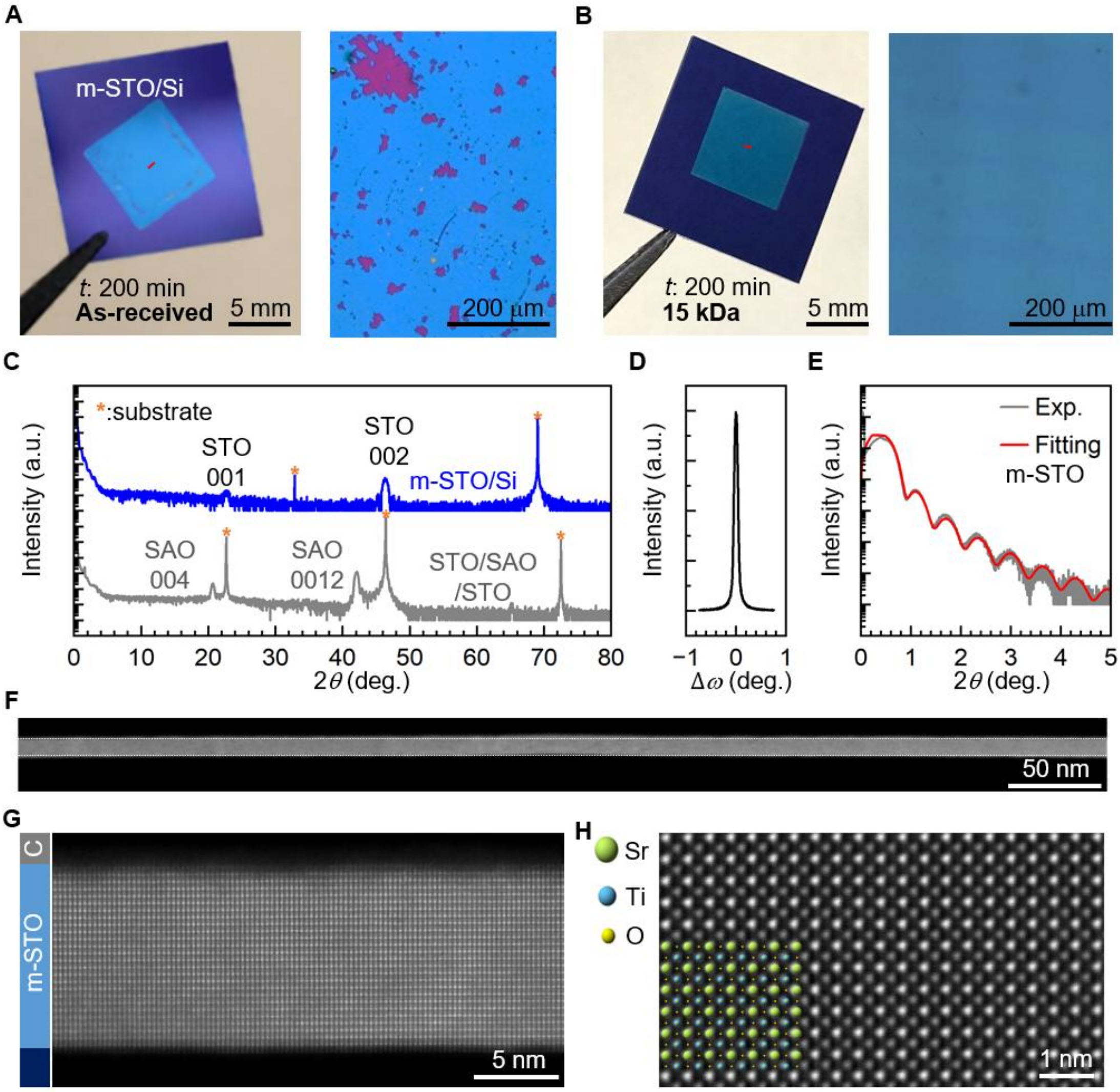

**Fig. 3. Structural integrity and characterization of freestanding STO membranes. (A, B)** Optical microscopy images of STO membranes transferred onto Si substrates after 200 min of water immersion using **(A)** as-received PPC and **(B)** 15 kDa PPC as the supporting polymer. The membrane transferred using as-received PPC exhibits extensive fragmentation, whereas the membrane supported by 15 kDa PPC remains macroscopically intact and shows a uniform surface over a large area. **(C)** X-ray diffraction (XRD) scans of the epitaxial STO heterostructure and freestanding STO membranes transferred onto (001)-oriented Si substrate. Only (00*l*) reflections of STO are observed. Peaks marked with asterisks originate from the corresponding substrates. **(D)** Rocking-curve scan of the STO 002 reflection, indicating high crystalline quality of the transferred membrane. **(E)** X-ray reflectivity (XRR) profile and the corresponding fit, confirming the uniform thickness and smooth interfaces of the membrane. Cross-sectional **(F)** low- and **(G)** high-magnification scanning transmission electron microscopy (STEM) images of a transferred STO membrane, showing a sharp and flat STO (blue)/$SiO_2$ (purple) interface, indicating flatness of the STO membrane. **(H)** Atomic resolution plane-view HAADF image resolving well-ordered perovskite lattices of a freestanding STO membrane.

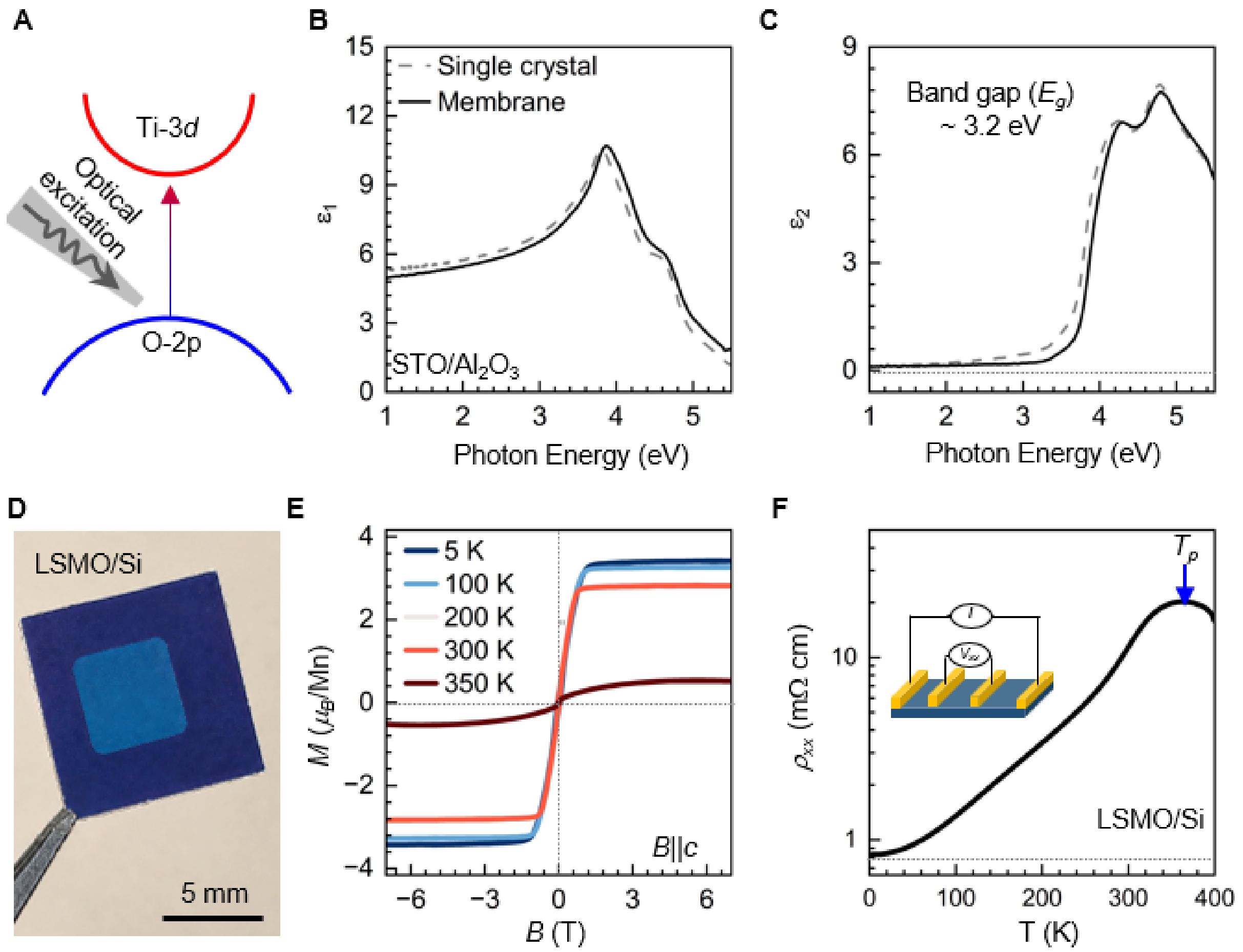

**Fig. 4. Optical, magnetic, and transport properties of transferred oxide membranes. (A)** Schematic illustration of the O 2*p* → Ti 3*d* interband transition responsible for the optical response of STO above 3.2 eV. **(B)** Real ($\varepsilon_1$) and **(C)** imaginary ($\varepsilon_2$) parts of the dielectric function for the freestanding STO membrane (black solid lines) and STO single crystal (grey dashed lines). The membrane exhibits optical responses closely matching those of bulk STO, indicating that the transfer process preserves the intrinsic electronic structure and optical properties. **(D)** Optical image of transferred $La_{0.7}Sr_{0.3}MnO_3$ (LSMO) on Si. **(E)** Magnetic hysteresis loops of the LSMO/Si membrane measured at temperatures from 5 to 350 K with the magnetic field applied along the out-of-plane direction (B||c). (F) Temperature dependence of the longitudinal resistivity, $\rho_{xx}$, of the LSMO/Si membrane, showing metallic behavior at low temperatures and a resistivity maximum at $T_p \sim 363$ K, associated with the metal–insulator crossover.